# PERFORMANCE EVALUATION OF IMPULSE RADIO ULTRA WIDE BAND WIRELESS SENSOR NETWORKS

Aubin
Lecointre[1, 2]
alecoint@laas.fr

Abdoulaye
Berthe[1, 2]
aberthe@laas.fr

Daniela
Dragomirescu[1, 2]
daniela@laas.fr

Jacques
Turbert[3]
jacques.turbert@dga.
defense.gouv.fr

Robert.
Plana[1, 2]
plana@laas.fr

[1] CNRS; LAAS; 7 avenue du colonel Roche, F-31077 Toulouse, France
[2] University of Toulouse; UPS, INSA, INP, ISAE; LAAS; F-31000 Toulouse, France
[3] Direction générale de l'armement CELAR/TEC La Roche Marguerite BP 57519 35174 BRUZ CEDEX France.

## ABSTRACT

*This paper presents a performance evaluation of Wireless Sensor Networks (WSN) based on Impulse Radio Ultra Wideband (IR-UWB) over a new simulation platform developed for this purpose. The simulation platform is built on an existing network simulator: Global Mobile Information System Simulator (GloMoSim). It mainly focuses on the accurately modeling of IR-UWB PHYsical (PHY) and Medium Access Control (MAC) layer. Pulse collision is modeled according to the used time hopping sequence (THS) and the pulse propagation delay in order to increase the simulation fidelity. It also includes a detection and identification application based on a new sensing channel and new sensor device models. The proposed architecture is generic so it can be reused for any simulation platform. The performance evaluation is based on one of the typical WSN applications: local area protection, where sensor nodes are densely scattered in an access regulated area in order to detect, identify and report non authorized accesses to a base station for analysis. Two networks topologies using different protocol stacks are investigated. Their performance evaluation is presented in terms of reliability and latency.*

## 1. INTRODUCTION

Wireless Sensor Networks can be defined as systems composed of several autonomous nodes linked together by a dedicated wireless link [1]. The nodes architecture may include a microprocessor, several sensor and actuator modules and also a radio communication module on a single board. WSNs support a large range of applications: monitoring, local area control, factory and house automation and tactical applications [1-3]. The case study presented in this paper studies a local area protection system. It is a kind of remote detection and identification application, in which sensor nodes are densely scattered in the protected area to detect or sense intrusion events, generated by intruder nodes presence in their vicinity, in order to report it to a base station for analysis. This can be used to reinforce homeland or military troop's security in a tactical application. The intrinsic constraints when setting up such systems are power efficiency, reliability, latency, simplicity, and small size [1-3]. IR-UWB is a good candidate to satisfy the mentioned constraints because of its interesting characteristics which are low radiated power, simple circuitry, localization ability, high multipath resolution and multiuser access capabilities using Time Hopping (TH) [4-5].

The goal of this paper is to analyze and propose an efficient WSN architecture based on IR-UWB and validate it using engineering simulation. As an alternative MAC-PHY, layer for 802.15.4a based WSN, several IR-UWB MAC-PHY models have been proposed [6-11]. These models can be divided into two categories: the first one insists on the PHY layer characterization [6-8]. The second one integrates this characterization into the network simulator [9-12]. None of them uses the real pulse propagation delay. Instead, they use a uniformly distributed random value to approximate it. This can be tolerated for the first type of models as they aim to provide a Bit Error Rate versus Signal and Interference to Noise Ratio (BER/SINR) depending on the number of active users. However, when modeling at the network simulator, such approximation can be avoided, as the pulse propagation delay and the number of active users is available.

Indeed, the second type of model does not completely meet the WSN simulation requirements as it does not include sensing and sensor channel models. This paper presents an overview of a new developed simulation platform for IR-UWB that takes into account the previous mentioned aspects. It also presents a comprehensive performance evaluation of WSNs that has been conducted using this platform. The performance evaluation compares distributed MAC protocol for IR-UWB to 802.15.4 Uncoordinated Access. The network performance is evaluated using a detection and identification application and also Constant Bit Rate (CBR) traffic. CBR is included for comparison purposes as it is mainly the used model to simulate WSN traffic.

The remainder of this paper will be organized as follows. Section 2 gives an overview of the developed simulation platform. Section 3 presents the performance evaluation scenario and their numerical analysis results analysis and finally Section 4 concludes.

## 2. SIMULATION PLATFORM OVERVIEW

We developed a WSN simulator based on IR-UWB in our previous work [13]. The platform development is based on a hardware prototype [5]. It mainly focuses on the IR-UWB PHY and MAC layer accuracy modeling. The PHY layer behavior is modeled by taking into account the pulse collision according to the pulse propagation delay. Slotted and UnSlotted MAC protocols for IR-UWB are modeled. A remote detection and identification application is also involved.

### 2.1. Physical Layer Model

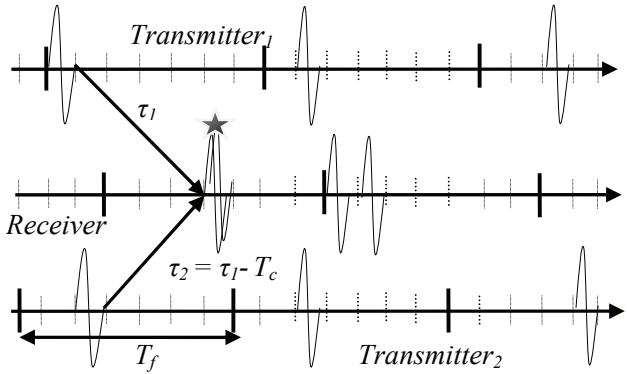

**Figure 1: Collision illustration**

IR-UWB signals are transmitted in form of very short pulses with low duty cycle (figure 1). The medium is divided into frames and each frame is shared in $N_h$ chips. The frame and chip duration are $T_f$ and $T_c$, respectively. The transmitted symbol can be repeated following a pseudo random sequence to avoid catastrophic collision under multiuser access conditions [7-8]. Using the Time Hopping Binary Pulse Amplitude Modulation (TH-BPAM) scheme for example, the $k^{th}$ user transmitted signal $s_{tx}^{(k)}(t)$ can be expressed as [7-8]

$$s_{tx}^{(k)}(t) = \sum_{j=-\infty}^{+\infty} \sqrt{E_{tx}}\, x_{tx}\left(t - j.T_f - c_j^k.T_c\right), \qquad (1)$$

where $E_{tx}$ is the transmitted pulse energy; $x_{tx}(t)$ denotes the basic pulse shape and $\left\{c_j^k\right\}$ represents the $j^{th}$ component of the pseudo random Time Hopping Sequence. The received signal $r(t)$ when only one user is present can be expressed as

$$r(t) = A.S_{tx}(t - \tau) + n(t), \qquad (2)$$

$$r(t) = \sum_{j=-\infty}^{+\infty} A.\sqrt{E_{tx}}\, .x_{tx}\left(t - j.T_f - c_j^k.T_c - \tau\right) + n(t), \qquad (3)$$

where $\tau$ represents the pulse propagation delay and $n(t)$ is Additive White Gaussian Noise (AWGN) with $N_0/2$ power density and $A$ represents the attenuation the signal experiences during propagation [7-8]. It depends on the considered channel model in terms of path loss, multipath, shadowing.

In a multi user scenario where $N_u$ users are active, the received signal is expressed as

$$r(t) = \sum_{k=1}^{k=N_u} A_k.S_{tx}(t - \tau_k) + n(t), \qquad (4)$$

$$r(t) = A_1.S_{tx}(t - \tau_1) + \sum_{k=2}^{N_u} A_k.S_{tx}(t - \tau_k) + n(t), \qquad (5)$$

where $\tau_k$ represents the delay associated to the propagation and asynchronism between clocks [7-8]. $A_k$ represents the attenuation of the $k^{th}$ user's signal (k=1 represents the signal of the user interest). This formulation can be used to characterize the TH-IR-UWB PHY layer in a multi user scenario and directly reports to the network simulator [9-12]; however the used propagation delay does not represent the real propagation delay for the real deployment configuration. The used Bit Error Rate (BER) versus the Signal to Interference and Noise Ratio (SINR) is also based on a perfect power control assumption which is not always realistic.

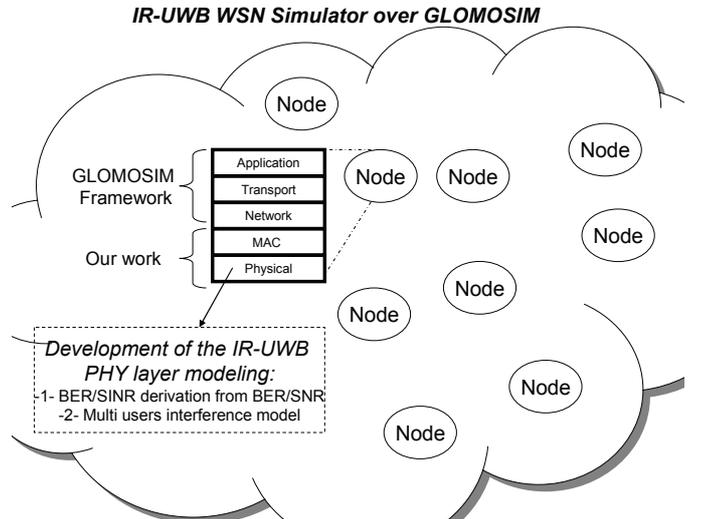

**Figure 2: Simulation Methodology Overview**

Instead of characterizing BER versus SINR of concurrent transmissions out of the network simulator in a multi user scenario and report it on the network simulator, our model is based on a two steps characterization process. We first perform an extensive Matlab/Simulink© simulation to obtain the relationship between the BER and the SNR: $E_b/N_0$ in a single user scenario. The BER versus SNR for IR-UWB can also be derived from point to point link measurement in the targeted environment.

The multi user interference characterization is reported to the network simulator PHY layer model for more accuracy. This constitutes the second characterization step in our model (figure 2). In this step we model the pulse interference according to the pulses' real propagation delay, during the concurrent transmission, instead of using Gaussian approximation to emulate the multi user interference. Indeed, Gaussian approximation to evaluate multi user interference has been proven to be unrealistic [8]. Moreover, our new scheme avoids an a priori assumption about the propagation delay $\tau_k$, the number of active users $N_u$ and the perfect power control ability as they are available during the simulation. The propagation delay is computed according to the node position, the pulse velocity and the occupied bandwidth [13]. The number of active users depends on the number of concurrent transmission being performed. The received power is evaluated according to the used channel model (Free Space, Rice or Rayleigh channel model).

The multiuser access interference is computed and added to the receiver background noise $n(t)$ on a chip per chip basis. This technique outperforms the model proposed in [9] in terms of accuracy. Indeed, in [9], the pulse propagation delay of concurrent transmission using the same or different THS is mainly modeled at the first characterization stage using a Gaussian approximation [8]. Note that the reception THS at a particular receiver depends on its local view of the medium frame structure (Figure 1). So it may vary depending on the node position and the central frequency of the occupied bandwidth. The $j^{th}$ component of the reception time hopping sequence $\rho_j^k$ of the $k^{th}$ user at a particular receiver can be expressed as

$$\rho_j^k = \left(T_c c_j^k + \tau_k\right) \bmod T_f. \qquad (6)$$

The reception THSs are computed and stored in an interference matrix $M$ (Figure 3). We use an interference vector $S$ to store the SINR of the signal pulses of the user of interest. For each received pulse, the SINR is dynamically updated.

The pulses that interfere with the user of interest ($user_I$) are the reception sequence $j^{th}$ elements defined by the interfering matrix content such as:

$$\left(T_c.c_j^1 + \tau_1\right) \bmod T_f = \left(T_c.c_j^k + \tau_k\right) \bmod T_{f.} \qquad (7)$$

$$\Leftrightarrow \rho_j^1 = \rho_j^k \qquad (8)$$

Doing the parallel between the previous equations and the received power $P_k$ of the concurrent reception, the received signal for the user of interest can be expressed as

$$P_{rx} = P_1 + \sum_{k=2}^{N_u} \left(\bar{P}_k + \frac{N_0}{2}\right), \qquad (9)$$

where $\bar{P}_k$ represents the received power of pulses located in the same frame.

$$\bar{P}_k = \begin{cases} P_k, \text{ if } \rho_j^1 = \rho_j^k \\ 0, \text{ otherwise} \end{cases} \qquad (10)$$

So the SINR vector S can be obtained as follow where the $j^{th}$ component is defined as :

$$S_j = \frac{P_1}{\dfrac{N_0}{2} + \displaystyle\sum_{k=2}^{N_u} \bar{P}_k}. \qquad (11)$$

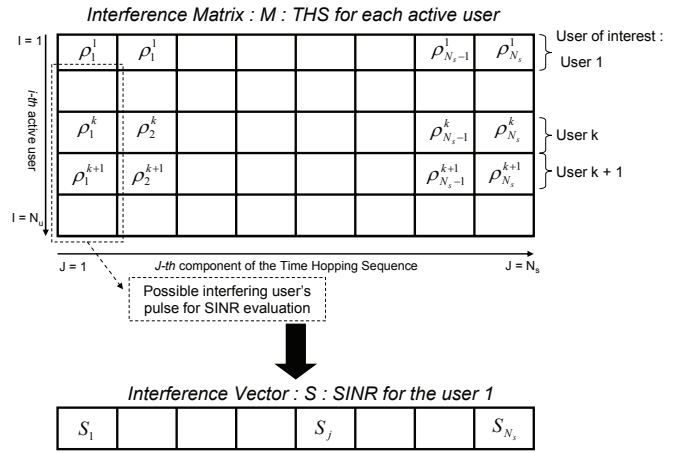

**Figure 3: Multi user interference illustration using an interference matrix**

This model is based on a single user reception model. However, multi user reception is possible once the preambles are well acquired, which means that the reception THS do not interfere. In this particular case the SINR vector $S$ has to be replaced by an SINR matrix as we are interested in decoding every receiving signal.

The presented methodology is generic, thus it can be used for any multiuser access scheme: Frequency Hopping Spread Spectrum (FHSS) as well as Direct Sequence Spread Spectrum (DSSS) for example.

## 2.2. MAC layer model

We modeled distributed Medium Access Control protocols for IR-UWB [14]: UnSlotted and Slotted MAC model. These are simple ALOHA [3] [15] like protocols with parameterized reliability and slot size. Their performances are evaluated and presented in the Section 3.

### 2.3 Sensor and sensing channel model

Detailed modeling of the sensor device is a key feature to obtain an accurate WSN simulation framework, as it has an impact on the network performance [16-17]. Our model is based on mechanic wave propagation. To set it up, we first characterize the sensor device and sensing channel by considering their important parameters: sampling rate, sensing range, missed detection rate. We use this characterization to mimic the sensor node behavior on the network simulator.

- The sensing range is modeled using a probabilistic detection range instead of full disc coverage.
- The signal propagation is modeled by a two ray ground reflection path loss and a Ricean fading multipath channel model.
- Missed detections are modeled using adjustable parameters.

The principle is summarized as follows: The targeted nodes periodically generate a signal at the sampling rate of the sensor device. This signal is sensed by the sensor node. According to its sensitivity, it detects or not the presence of an intruder.

The two defined thresholds represent the device sensitivity and its detection threshold for correct detection (figure 4). Furthermore, the signal generated by two or more targeted nodes may collide at the sensor device input, thus leading to missed detection. The presence of an intruder or a targeted node may not always be notified by the sensor device because of the additional attenuation due to multipath losses, thus leading to missed detection.

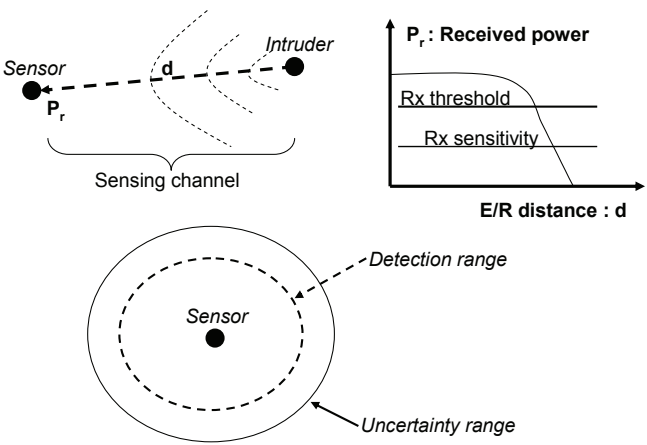

**Figure 4: Sensor and sensing channel**

This generic method can be used to represent many kind of sensor device behavior, after adjusting the mentioned parameters.

An example of a sensor device which can be modeled following the mentioned technique is a binary acoustic sensor present in the Mica Mote hardware. This kind of device provides one bit information regarding the presence or absence of an intruder node in its vicinity without 100% reliability [18].

### 3. PERFORMANCE EVALUATION

In this section, we present the performance evaluation of an example of WSN applications, the local area protection application, which is just a case study as our proposed architecture is generic and reusable. Remote detection and identification performance is evaluated in the context of low cost and low power WSN architectures. The configurations used at the MAC and PHY layer are: Carrier Sense Multiple Access with Collision Avoidance (CSMA/CA) MAC layer over an Offset Quadrature Phase Shift Keying (OQPSK) PHY and UnSlotted and Slotted MAC protocols over an IR-UWB PHY. The presented simulation results are based on a Uniform Pulse Train Spacing multi user access [7]. The relevant simulation parameters are summarized in Table 1. Two models are mainly considered.

**Table 1: Simulation parameters**

| Parameter | TH-IR-UWB | OQPSK |
|---|---|---|
| Bandwidth (MHz) | 100 | 2 |
| Frequency (GHz) | 0.8 | 2.45 |
| Throughput(Mbps) | 1 | 0.25 |
| Capture Model | *ber based* | *ber based* |
| Antenna Height (m) | 0.45 | 0.03 |
| Antenna Gain (dB) | 3 | 3 |
| Noise Figure (dB) | 5 | 10 |
| Temperature (K) | 270 | 270 |
| Sensitivity (dBm) | -85 | -96 |
| RX-Threshold (dBm) | -80 | -85 |
| TX-Power (dBm) | -24.318 | 17 |

### 3.1. First simulation scenario

The first one is a simple star topology in which CBR source to sink transmission is used to evaluate the network performances using static routing tables.

Figure 5 depicts the first simulation scenario in which four router nodes are placed around a base station. Four others nodes are placed in their vicinity to mimic the sensor nodes' behavior. Each of them generates CBR traffic. The network performance is evaluated under different traffic load condition.

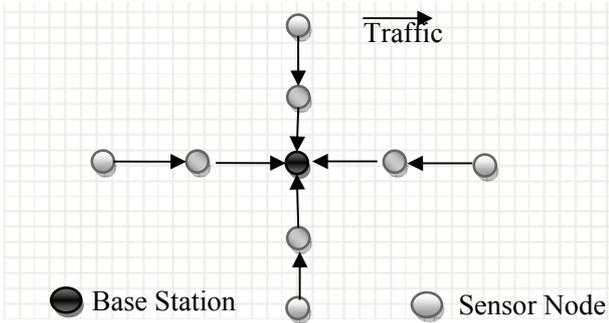

**Figure 5: Scenario 1 illustration**

The performance metrics evaluated in the first scenario are:

- The packet delivery ratio which expresses the ratio between the number of CBR application byte sent by the source and the number of received bytes at the destination.
- The average end to end delay which expresses the mean delay time from the source node to the destination.

In this section the simulation results are presented and analyzed. Figure 6 presents the variation of the packet delivery ratio as the number of retransmission increases from 0 to 6. Unexpectedly, the results show that 100% packet delivery ratio is obtained with UnSlotted protocol with 4 retransmissions whereas it is obtained in Slotted with 6 retransmissions.

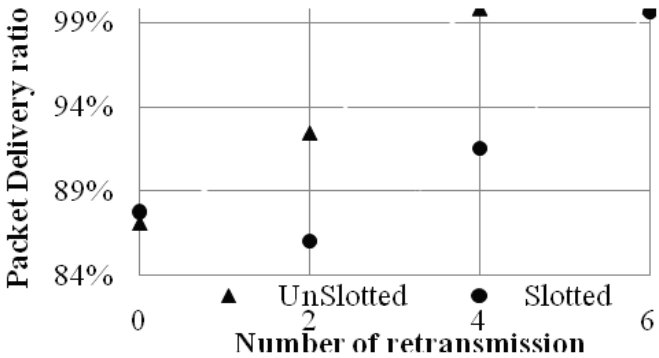

**Figure 6: Packet Delivery Ratio**

In fact, when employing the Slotted protocol, repeated code collisions seem often to occur because of the relative synchronization between nodes. Nodes always wait for the new slot front. This can be resolved by adding a random delay before starting the new transmission.

Figure 7 compares the average end to end delay as the number of retransmission increases in Slotted and UnSlotted protocol. It can be seen that the UnSlotted protocol has lower latency than the Slotted protocol. This is because when using the Slotted protocol, nodes must wait for the slot front before starting a transmission. The same experiment was conducted with the CSMA/CA over OQPSK without the Request To Send/Clear To Send RTS/CTS handshake. Here, the packet delivery ratio was 50.1% and the obtained average end to end delay was 7.58 E-03. This is mainly due to the losses induced by packet collision and the relatively low data rate (250kbps).

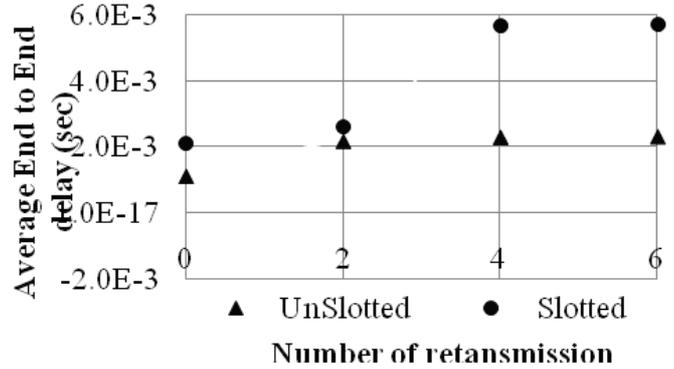

**Figure 7: Average End to End Delay**

In this first set of experiments, the traffic load has been varied from 0.1 packet/second to 80 packets per second. This did not affect the evaluated performance.

### 3.2. Second simulation scenario

The second model consists of a complete WSN system where sensor nodes are scattered in the protected area in order to detect, authenticate and track the intruder nodes. In this scenario, detection and authentication packets are sent to the sink node using a reactive multi hop ad hoc routing protocol: Ad hoc On Demand Distance Vector (AODV).

Figure 8 depicts the second scenario where 60 sensor nodes are placed around a base station to detect and eventually authenticate intruders. In this scenario, intruder nodes may be mobile, thus enabling tracking. Two types of nodes have been considered: Unauthorized and authorized nodes. Authorized nodes are able to respond to authentication request generated by the sensor devices.

The performance metrics evaluated in the second scenario are:

- The system reliability in terms of detection which expresses the ratio between the generated event and the notified event to the base station.
- The system reliability in terms of authentication which expresses the ratio between the authentication request and the notified responses to the base station.

- The detection latency which expresses the delay between an intrusion and its notification to the base station.

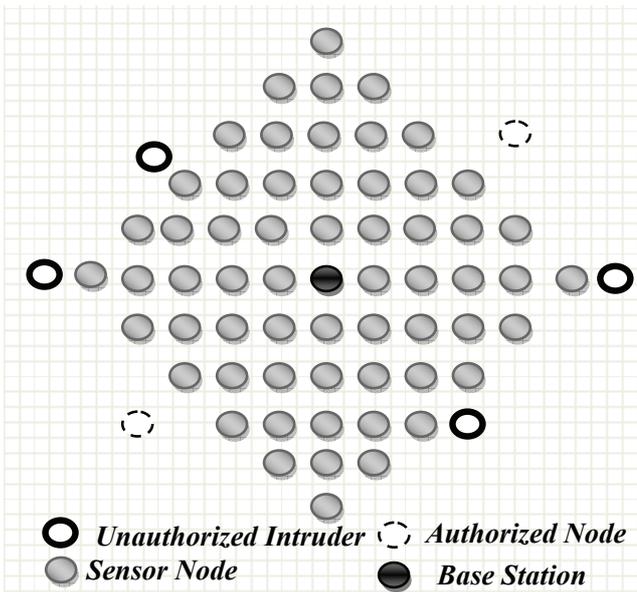

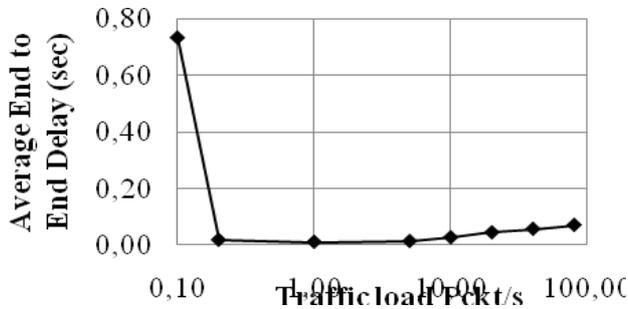

**Figure 8: Scenario 2 illustration**

The first experiment of the second scenario consists on a peer to peer CBR traffic with different traffic loads, five CBR applications are used between nodes located at different sides of the protected area.

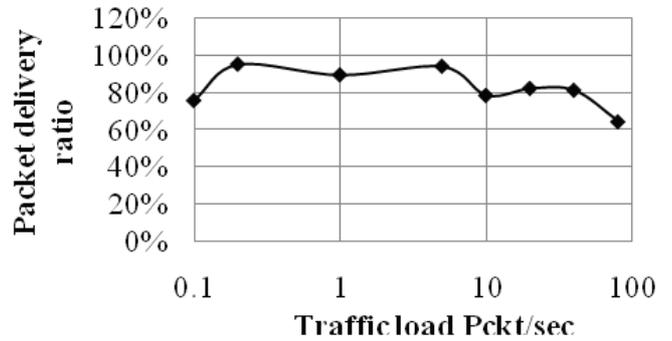

**Figure 9: Average End to End delay**

Figure 9 depicts the average end to end delay variation depending on the traffic load. The high value of the average end to end delay with 0.1 packets per second is due to the route establishment delay, caused by the routes TTL (Time To Live). In fact, in the AODV routing protocol, the routes need to be reconstructed if they last a certain time. So under light traffic conditions, almost every transmitted packet creates a route establishment overhead.

Figure 10 represents the packet delivery ratio in different traffic condition. As expected, the packet delivery ratio drops as the traffic grows. The low packet delivery ratio with 0.1 packets per second is also due to the routing protocol overhead in light traffic condition. With traffic load above 0.2 packets per second the effect of the routing protocol overhead disappears, as the established paths are frequently used.

**Figure 10: Packet Delivery Ratio**

Figure 11 and 12 are plotted using our developed sensing and sensor channel model with 95% reliability. Their variation is quite similar to the CBR traffic one. However, the detection and authentication rate which are linked to the packet delivery ratio are not the same. This demonstrates the inaccuracy of approximating WSN application with CBR traffic as already proven in [18].

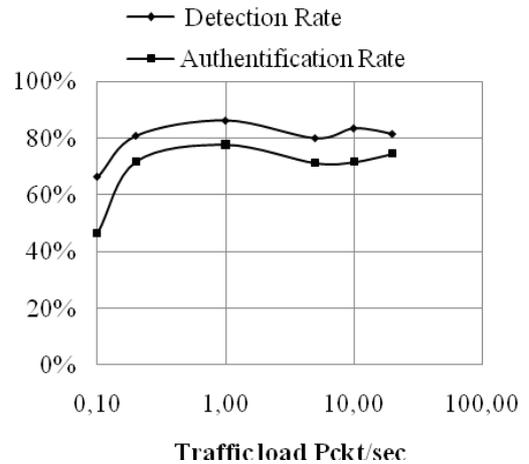

**Figure 11: Detection and Authentication Rate**

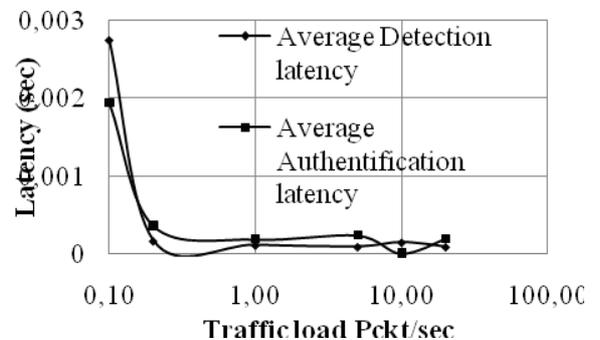

**Figure 12: Detection and Authentication Latency**

# 4. CONCLUSION

In this paper, we presented a WSN simulation architecture for TH-IR-UWB MAC and PHY layers. The proposed simulator accurately deals with the IR-UWB specificities. It proposes a new scheme to accurately model the multiuser access with Time Hopping Impulse Modulation on a network simulator. Furthermore, the proposed multi user interference modeling scheme can be reused for all Code Division Multiple Access (CDMA) techniques. This new scheme uses the time chip-scale division for evaluating in real time the SINR of the PHY link. For increasing accuracy the real pulse propagation delay is used. Thus it allows a more accurate multi user interference and pulse collision model.

The presented simulator includes sensor and sensing channel models based on mechanic wave propagation and a detection and identification application. This scheme has been implemented using the network simulator GloMoSim. Using the developed platform, several experiments have been conducted; they demonstrate the ability for IR-UWB to match the WSN constraints in term of reliability and latency. The performance evaluation shows that Unslotted MAC is more efficient than Slotted MAC for IR-UWB. It demonstrates that using Unslotted MAC with IR-UWB PHY is 50% more reliable and more latency efficient than the CSMA MAC for OQPSK PHY.

Our future work will include tracking algorithm performance evaluation based on IR-UWB positioning capabilities, as well as low cost and low power sensor node hardware architecture prototyping based on IR-UWB. In addition an improvement of the sensing an channel model has to be proposed for enabling not only binary sensor modeling. Thanks to the high scalability of GloMoSim, these IR-UWB PHY and MAC improvements can be used for WSN architecture evaluation and optimization.